\documentclass[12pt]{article}
\usepackage{amsmath,latexsym,amssymb,epsfig,psfrag}
\usepackage{axodraw}
\usepackage{graphicx}  
\newcommand{\BLambda}{\boldsymbol{\lambda}}
\def\ltsim{\lower3pt\hbox{$\, \buildrel < \over \sim \, $}}  
\def\gtsim{\lower3pt\hbox{$\, \buildrel > \over \sim \, $}}  
\newcommand{\be}{\begin{equation}}  
\newcommand{\ee}{\end{equation}}  
\def\ga{\mathrel{\raise.3ex\hbox{$>$\kern-.75em\lower1ex\hbox{$\sim$}}}}  
\def\la{\mathrel{\raise.3ex\hbox{$<$\kern-.75em\lower1ex\hbox{$\sim$}}}}  

\makeatletter   
\@addtoreset{equation}{section}   
\makeatother

\def\simlt{\stackrel{<}{{}_\sim}}
\def\simgt{\stackrel{>}{{}_\sim}}  

\begin{document}
\baselineskip=16pt   

\begin{titlepage}  

\vskip 2cm
\begin{flushright}
{\bf CERN-PH-TH/2006-188}\\
{\bf LPTHE-06-04}\\
{\bf UAB-FT-611} 
\end{flushright}
\begin{center}  
\vspace{0.5cm} \Large {\sc 
A New Gauge Mediation Theory}
\vspace*{5mm}  
\normalsize  
  
{\bf
I.~Antoniadis~$^{a,}$\footnote{On leave of absence from CPHT, Ecole
Polytechnique, UMR du CNRS 7644, F-91128 Palaiseau Cedex},
K.~Benakli~$^{b}$, 
A.~Delgado~$^{a}$
and
M.~Quir\'os~$^{c}$
}  

\smallskip   
\medskip   
\it{$^{a}$~Department of Physics, CERN -- Theory Division, 1211 Geneva
23, Switzerland}\\

\smallskip   
\medskip 
\it{$^{b}$~Laboratoire de Physique Th\'eorique et Hautes
Energies}\\ 
\it{Universit\'es de Paris VI et VII, France}

\smallskip   
\medskip 
\it{$^{c}$~Instituci\'o Catalana de Recerca i Estudis
Avan\c{c}ats (ICREA)}\\ {\it and}\\
\it{Theoretical Physics
Group, IFAE/UAB, E-08193 Bellaterra, Barcelona, Spain}

\vskip0.6in 
\end{center}  
   
\centerline{\large\bf Abstract}  

\noindent
We propose a class of models with gauge mediation of supersymmetry
breaking, inspired by simple brane constructions, where $R$-symmetry
is very weakly broken. The gauge sector has an extended $N=2$
supersymmetry and the two electroweak Higgses form an $N=2$
hypermultiplet, while quarks and leptons remain in $N=1$ chiral
multiplets. Supersymmetry is broken via the $D$-term expectation value
of a {\it secluded} $U(1)$ and it is transmitted to the Standard Model
via gauge interactions of messengers in $N=2$ hypermultiplets:
gauginos thus receive Dirac masses.  The model has several distinct
experimental signatures with respect to ordinary models of gauge or
gravity mediation realizations of the Minimal Supersymmetric Standard
Model (MSSM). First, it predicts extra states as a \emph{third}
chargino that can be observed at collider experiments. Second, the
absence of a $D$-flat direction in the Higgs sector implies a lightest
Higgs behaving exactly as the Standard Model one and thus a reduction
of the `little' fine-tuning in the low $\tan\beta$ region.  This
breaking of supersymmetry can be easily implemented in string theory
models.
   
\vspace*{2mm}   
%\smallskip\newline  
  
\end{titlepage}  
  
\section{\sc Introduction}
\label{introduction}

The supersymmetric flavor problem is one of the guidelines for
constructing realistic models at the electroweak scale with deep
implications at LHC. Supersymmetry is generically broken in a hidden
sector and transmitted to the observable sector either by gravity or
by gauge interactions giving rise to the so-called
gravity~\cite{Nilles:1983ge} or gauge-mediated~\cite{Giudice:1998bp}
models. In gravity-mediated scenarios the soft-breaking masses are
generated at the Planck scale and there is no symmetry reason why they
should be flavor-invariant or why they should not mediate large
contributions to flavor-changing neutral current processes. In fact
this idea motivated the introduction of gauge-mediated theories where
the generated soft terms are flavor-blind and therefore they feel
flavor breaking only through Yukawa interactions.

In gauge mediated (GMSB) theories supersymmetry is broken in a {\it
secluded} sector such that the Goldstino field belongs to a chiral
superfield $X$ which acquires a vacuum expectation value (VEV) along
its auxiliary $F$-component as $\langle X\rangle=M+\theta^2 F$, where
it is usually assumed that $F\ll M^2$ so that supersymmetry breaking
can be treated as a small perturbation.  The theory also contains a
vector-like {\it messenger} sector $(\Phi,\Phi^c)$ coupled to $X$ by a
superpotential coupling $\int d^2\theta \Phi^c X\Phi$ and with
Standard Model (SM) quantum numbers providing one-loop Majorana masses
to gauginos $\sim \alpha/4\pi\,F/M$ and two-loop positive squared
masses to all sfermions as $\sim (\alpha/4\pi)^2\, F^2/M^2$, where
$\alpha=g^2/4\pi$ and $g$ is the corresponding gauge coupling. The
corresponding effective operators giving rise to these masses are
$(1/M)\int d^2\theta X {\rm Tr} W^\alpha W_\alpha$ and $(1/M^2)\int
d^4\theta X^\dagger X \mathcal Q^\dagger \mathcal Q$, where $W^\alpha$
are the chiral field strengths of the SM vector superfields and
$\mathcal Q$ the SM chiral superfields. Fixing all masses in the TeV
range one obtains a relation between $F$ and $M$. In particular if
gravity mediated contributions are required not to reintroduce the
flavor problem, the flavor changing Planck suppressed contributions
should remain smaller than the flavor conserving gauge-mediated ones,
i.e.~$M\simlt (\alpha/4\pi)\, M_{P\ell}\sim 10^{16}$ GeV, implying
$\sqrt{F}\simlt 10^{11}$ GeV. In gauge mediation the $R$-symmetry is
broken by the same mechanism that breaks supersymmetry which
constrains the corresponding ultraviolet (UV) completion of the
theory. The main problem of any fundamental theory, as e.~g.~string
theory, is then to provide the required mechanism of supersymmetry and
$R$-symmetry breaking.

In the simplest compactifications of a class of string theories,
leading to intersecting branes at angles, the gauge group sector is
often in multiplets of extended supersymmetry while matter states come
in $N=1$ multiplets~\cite{Uranga:2005wn}. A simple way of breaking
supersymmetry is by deforming the intersection angles from their
special values corresponding to the supersymmetric configuration. A
small deformation $\epsilon$ of these angles breaks supersymmetry via
a $D$-term vacuum expectation value, associated in the $T$-dual
picture to a magnetized $U(1)$ in the internal compactification space,
$\epsilon=D\ell_s^2$ with $\ell_s$ the string length. In this case,
the {\it secluded} sector is given by an (extended) vector multiplet
containing an $N=1$ vector superfield $V^0$.  The Goldstino is then
identified with the gaugino of $V^0$ which acquires a VEV: $\langle
V^0 \rangle =\frac{1}{2}\theta^2\bar\theta^2 D$ or $\langle
W^0_\alpha\rangle =\theta_\alpha D$.  Of course, this supersymmetry
breaking preserves the $R$-symmetry and can only give Dirac masses to
the gauginos of the extended gauge sector.

A prototype model is based on type II string compactifications on a
factorizable six-torus $T^6=\otimes_{i=1}^3 T_i^2$ with appropriate
orbifold and orientifold planes and two sets of brane stacks: the
observable set $\mathcal O$ and the hidden set $\mathcal
H$~\cite{Antoniadis:2005em,Antoniadis:2006eb}. The SM gauge sector
corresponds to open strings that propagate with both ends on the same
stack of branes that belong to $\mathcal O$: it has therefore an
extended $N=2$ or $N=4$ supersymmetry. Similarly, the secluded gauge
sector corresponds to strings with both ends on the hidden stack of
branes $\mathcal H$.  The SM quarks and leptons come from open strings
stretched between different stacks of branes in $\mathcal O$ that
intersect at fixed points of the three torii $T_i^2$ and have
therefore $N=1$ supersymmetry. The Higgs sector however corresponds to
strings stretched between different stacks of branes in $\mathcal O$
that intersect at fixed points of two torii and that are parallel
along the third one: it has therefore $N=2$ supersymmetry.  Finally
the messenger sector contains strings stretched between stacks of
branes in $\mathcal O$ and the hidden branes $\mathcal H$, that
intersect at fixed points of two torii and are parallel along a third
torus. It has therefore also $N=2$ supersymmetry. Moreover, the two
stacks of branes along the third torus are separated by a distance
$1/M$, which introduces a supersymmetric mass $M$ to the
hypermultiplet messengers. The latter are also charged under the
supersymmetry breaking $U(1)$ and they are given corresponding
supersymmetry breaking squared-masses $\pm D$.

In this paper (inspired by the previous brane constructions) we
propose a new gauge mediated theory (NGMSB), alternative to the usual
GMSB, where $R$-symmetry remains unbroken by the mechanism of
supersymmetry breaking~\footnote{As we will see later, gravitational
interactions produce $R$-symmetry breaking at a subleading level in
the observable sector.}.  The observable gauge sector is a set of
$N=2$ gauge multiplets corresponding to the SM gauge group. In general
an $N=2$ gauge multiplet contains an $N=1$ vector multiplet
$V=(A_\mu,\lambda_1,D)$ and a chiral multiplet
$\chi=(\Sigma,\lambda_2,F_\chi)$. It can be described by the $N=2$
chiral vector superfield~\cite{Alvarez-Gaume:1996mv}
\begin{equation}
\mathbb A=\chi+{\tilde\theta}W+{\tilde\theta}^2{\mathcal D}{\mathcal D}\chi\ ,
\label{supercampo2}
\end{equation}
where $\tilde\theta$ is the second $N=2$ Grassmannian coordinate and
${\mathcal D}$ the $N=1$ super-covariant derivative.  The secluded
sector is identified with the $N=2$ chiral vector superfield $\mathbb
A^0$ (whose $N=1$ vector superfield is $V^0$) where supersymmetry is
broken by a hidden $D$-term as~\footnote{We are assuming here that
some dynamical mechanism in the hidden sector generates the
$D$-breaking of supersymmetry given by the VEV in
Eq.~(\ref{secluded}). A particular realization is given by the angle
deformation of intersecting branes described above, leading to an
$N=2$ Fayet-Iliopoulos term.}
\begin{equation}
\langle\mathbb A^0\rangle =\tilde\theta\langle
W^0\rangle={\tilde\theta}\theta D
\label{secluded}
\end{equation}
For the messenger sector, we choose a (set of) $N=2$ hypermultiplet(s)
denoted as $(\Phi^c,\Phi)$, with field contents
$\Phi=(\phi,\psi,F_\Phi)$ and $\Phi^c=(\phi^c,\psi^c,F_{\Phi^c})$,
charged under the secluded $U(1)$ (with charges $\pm 1$) and with a
supersymmetric mass $M$.

The content of the paper is as follows. In section~\ref{model} we
present the structure of our model. In particular the transmision of
supersymmetry breaking from the hidden to the observable sector is
calculated by loop diagrams involving messenger fields. Its main
outputs are Dirac masses for gauginos of the $N=2$ gauge sector
($R$-symmetry is preserved by the $D$-breaking) as well as soft masses
for $N=1$ sfermions. In section~\ref{electroweak} we study the
generation of the soft-breaking terms for the $N=2$ Higgs sector, as
well as the electroweak symmetry breaking. The main departure of the
latter from the MSSM is the absence of $D$-flat directions along which
the Higgs potential becomes unstable. As a consequence, after
electroweak symmetry breaking, the SM-like Higgs has a
$\tan\beta$-independent mass and couplings to SM fermions while the
rest of the Higgs sector has $\tan\beta$-independent masses (at the
tree-level) and $\tan\beta$-dependent couplings to ordinary matter. In
section~\ref{phenomenology} we present a few comments about the
gravitino mass and the generation of an $F$-breaking by gravitational
interactions, possible dark matter candidates in our scenario, its
experimental signatures at the next high energy colliders (LHC and
ILC) and the possibility of unification at the GUT scale. In
section~\ref{conclusion} our conclusions are drawn. Finally in
appendix A we give a description of the supersymmetry breaking
potential, based on a Fayet-Iliopoulos (FI) term, for the scalar
messengers and the scalar field in the chiral multiplet in the
secluded gauge sector. We show that the supersymmetry $D$-breaking
minimum is a local (metastable) one. However we also show that it is
absolutely stable on cosmological times.

\section{\sc The model structure}
\label{model}
The interaction Lagrangian of the messengers with the different gauge
sectors is written as
\begin{equation}
\int d^4\theta \left\{ \Phi^\dagger e^{V}\Phi+\Phi^c
e^{-V}\Phi^{c\dagger}\right\}+\left\{\int d^2\theta 
\Phi^c\left[M-\sqrt{2}\;\chi \right]\Phi+h.c.\right\}
\label{lag1}
\end{equation}
where $V=\sum_A g_A T^A V^A$ contains all gauge fields (in the hidden
and observable sectors) and similarly for $\chi=\sum_A g_A
T^A\chi^A$~\footnote{We are normalizing the generators such that ${\rm
Tr} T^A T^B=1/2\, \delta^{AB}$ in the fundamental representation.}.

>From (\ref{lag1}) and after replacing the $V^0$ VEV from
Eq.~(\ref{secluded}), we find that the Dirac spinor
$\Psi=(\psi,\bar\psi^c)^T$ acquires a Dirac mass $M$ while the scalar
components have a squared mass matrix given by
\begin{equation} (\phi^\dagger,\phi^c)\mathcal M^2\left(
\begin{array}{c}
\phi\\
\phi^{c\dagger}
\end{array}
\right),\quad
\mathcal M^2=\left(
\begin{array}{cc}
M^2+D& 0\\
0& M^2-D
\end{array}
\right)
\label{mass0}
\end{equation}
Notice that in the absence of the supersymmetric mass $M$, the origin
in the $(\phi,\phi^c)$ field space is a saddle point. However since we
are assuming that $M^2> D$ the origin becomes a minimum along the
$(\phi,\phi^c)$ directions. Assuming that supersymmetry is broken by a
FI mechanism the $\Sigma^ 0$-scalar in the extended gauge sector is a
flat direction of the potential.  However in the presence of
supersymmetry breaking it will acquire a radiative soft mass, see
Eq.~(\ref{mSigma}), and the origin
$\langle\Sigma^0\rangle=\langle\phi\rangle=\langle\phi^c\rangle=0$
becomes a local minimum.  Of course there exists a global minimum
where supersymmetry is restored and gauge symmetry is broken, at
$\langle\Sigma^0\rangle=M/\sqrt{2}$ and $|\langle\phi^c\rangle|^2=D$.
The barrier heigh separating these minima is very small as compared to
the distance between them if $M^2\gg D$. In that case the tunneling
probability per unit space-time volume from the local minimum to the
global one is
\begin{equation}
\mathcal P\sim e^{-{\displaystyle \kappa\,M^4/D^2}}
\label{tunnel}
\end{equation}
where $\kappa> 1$ is a dimensionless constant. For $M^2\gg D$ the
probability (\ref{tunnel}) is so small that the false vacuum is
essentially stable on cosmological times. More details can be found in
appendix A.

Using now the Lagrangian (\ref{lag1}), the coupling of the SM gauginos
with the messenger fields is written as
\begin{equation}
-\sqrt{2}\left( \phi^c \bar\BLambda_1
 \Psi-\phi^\dagger\bar\BLambda_2\Psi\right)+h.c.
\label{lag2}
\end{equation}
where we have defined the four-component symplectic-Majorana spinors
$\BLambda_i=(\lambda_i,\epsilon_{ij}\bar\lambda_j)^T$~\footnote{Notice that if
we define the Dirac spinor $\BLambda=(\lambda_1,\bar\lambda_2)^T$, it
satisfies the identity
$
\bar\BLambda\BLambda=\frac{1}{2}\left(\bar\BLambda_1\BLambda_1-
\bar\BLambda_2\BLambda_2\right)\, .
$
}. A Dirac mass $m^D_a$ for the Dirac gaugino $\BLambda_a$ is
radiatively generated from the diagram of Fig.~\ref{gauginomass} which
gives a finite value
\begin{figure}
\begin{center}\begin{picture}(300,56)(0,0)%\SetScale{2}
%\Vertex(180,10){1.5}\Vertex(120,10){1.5}
\Line(100,10)(200,10)
\DashCArc(150,10)(30,0,180){3}
\Text(150,0)[]{$\Psi$}
\Text(150,50)[]{$\phi,\phi^c$}
\Text(110,0)[]{$\lambda_1$}\Text(200,0)[]{$\lambda_2$}
\end{picture}\end{center}
\caption{\it Feynman diagrams contributing to the Dirac mass of the
gaugino $\BLambda^a$}
\label{gauginomass}
\end{figure}
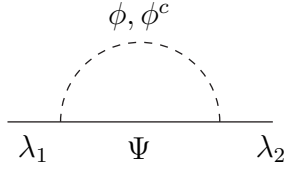
\begin{equation}
m^D_a=k_a\frac{\alpha_a}{4\pi}{\mathcal N}\frac{D}{M}\left[1+\mathcal
O\left(\frac{D^2}{M^4}\right)\right],\quad a=1,2,3
\label{mD}
\end{equation}
with $k_1=5/3$, $k_3=k_2=1$, $\alpha_a=g^2_a/4\pi$, and ${\mathcal N}$
the number of messengers.  This value for the Dirac gaugino mass can
be equivalently understood from the effective
operator~\cite{Fox:2002bu,Antoniadis:2005em}
\begin{equation}
\sim\frac{1}{M}\int d^2\theta W^0 {\rm Tr}(W\chi)+h.c.
\label{effec}
\end{equation}
This operator is actually consistent with $N=2$ supersymmetry since it
is generated by a manifest $N=2$ supersymmetric
Lagrangian~(\ref{lag1}).  The operator~(\ref{effec}) can be rewritten
in an explicit $N=2$ supersymmetric way:
\begin{equation}
\sim\frac{1}{M}\int d^2\theta d^2{\tilde\theta} \mathbb A^0 {\rm Tr}
(\mathbb A)^2+h.c.
\label{effec2}
\end{equation}
where $\mathbb A^0$ is the {\it secluded} $U(1)$ $N=2$ vector
superfield.

In Ref.~\cite{Antoniadis:2006eb} such an operator was computed in
string theory for the same physical setup of brane configurations
discussed in section~\ref{introduction}.  The result was found to be
topological, in the sense that it is independent of the massive string
oscillator modes. It receives contributions only from the field theory
Kaluza-Klein (KK) part of the torus along which the messengers brane
intersection of the observable stack $\mathcal O$ with the hidden
stack $\mathcal H$ is extended. The separation $\ell$ of the two
stacks along a direction within this torus determines the messengers
mass $M=\ell$ in string units. The gaugino mass (in the limit
$D\ell_s^2\ll 1$) can then be written as an integral over the real
modulus parameter $t$ of the worldsheet annulus having as boundaries
the two brane stacks~\footnote{Note a factor of $t$ misprint in
Eq.~(3.22) of Ref.~\cite{Antoniadis:2006eb}. Here, we also restored
the numerical prefactor.}:
\begin{equation}
m_{1/2}^D\simeq \frac{\alpha}{2}{\mathcal N}D \int_0^\infty
\sum_{n=-\infty}^{+\infty}(nR+\ell)e^{-2\pi t(nR+\ell)^2}\, ,
\label{mgstring}
\end{equation}
where for simplicity we have chosen the brane separation to be along
one of the two dimensions of an orthogonal torus of radius $R$.  The
integration can be performed explicitly with the result:
\begin{equation}
m_{1/2}^D\simeq \frac{\alpha}{4\pi}{\mathcal N}D\left\{ \frac{1}{\ell}
- \frac{2}{R^2}\sum_{n\ge
  1}\frac{\ell^2/R^2}{n^2-\ell^2/R^2}\right\}\, .
\label{mgstring2}
\end{equation}
The first term in the right-hand side reproduces precisely the field
theory expression (\ref{mD}), while the second term represents the
(sub-leading) contribution of the messengers KK excitations.

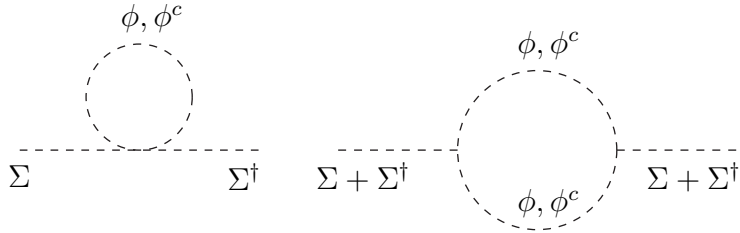
\begin{figure}[htb]\vspace{.5cm}
\begin{center}\begin{picture}(300,56)(0,0)%\SetScale{2}
%\Vertex(180,10){1.5}\Vertex(120,10){1.5}
\DashLine(30,10)(120,10){3}
\DashCArc(75,30)(20,0,360){3}
\Text(80,60)[]{$\phi,\phi^c$}
\Text(30,0)[]{$\Sigma$}\Text(115,0)[]{$\Sigma^\dagger$}
%
%\Vertex(180,10){1.5}\Vertex(120,10){1.5}
\DashLine(150,10)(195,10){3}
\DashLine(255,10)(300,10){3}
\DashCArc(225,10)(30,0,360){3}
\Text(230,50)[]{$\phi,\phi^c$}
\Text(230,-10)[]{$\phi,\phi^c$}
%\Text(225,90)[]{$\phi$}
%\Text(225,10)[]{$\phi^c$}
\Text(160,0)[]{$\Sigma+\Sigma^\dagger$}
\Text(285,0)[]{$\Sigma+\Sigma^\dagger
$}
\end{picture}\end{center}
\caption{\it Bosonic diagrams contributing to squared masses for the
adjoint scalars $\Sigma^a$}
\label{Sigmamass}
\end{figure}

The superpotential term in Eq.~(\ref{lag1}) gives rise to the $F$-term
potential
\begin{equation}
{\rm Tr}\left\{ 2\,\phi^\dagger \Sigma^\dagger\Sigma\phi+2\,\phi^c
\Sigma^\dagger\Sigma\phi^{c\dagger}-
\sqrt{2}M\left[\phi^\dagger(\Sigma+\Sigma^\dagger)\phi+
\phi^c(\Sigma+\Sigma^\dagger)\phi^{c\,\dagger}\right]\right\}\ .
\label{Fterm}
\end{equation}
By using now the one-loop diagrams of
Fig.~\ref{Sigmamass} we obtain for the real and imaginary parts
of the field $\Sigma^a$, a squared mass splitting
$m_{\Sigma_R^a}^2=-m_{\Sigma_I^a}^2\simeq {\displaystyle
k_a\frac{\alpha_a}{4\pi}\mathcal N \frac{D^2}{M^2}}$ that makes the
direction $\Sigma_I^a$ to be unstable~\footnote{We thank Yuri Shirman
for pointing out a mistake in an earlier version when computing the
$\Sigma^a$ squared mass.}.  This behaviour can be understood from the
effective operator
\begin{equation}
\sim \frac{1}{M^2}\int d^2\theta (W^0)^2 {\rm Tr}(\chi)^2 \ .
\label{effec3}
\end{equation}
This problem was discussed at length in Ref.~\cite{Chacko:2004mi}
where several solutions were proposed. A simple solution can be based
on the fact that, as we will see in section~4, cancellation of the
cosmological constant when supergravity interactions are turned on
will generically induce an $F$-breaking of the order of $D$ giving
rise to a spurion superfield $X=\theta^2 F$ in the hidden brane
$\mathcal H$. Interactions between $X$ in the hidden sector and $\chi$
can give rise through the operator
\begin{equation}
\sim\frac{1}{M^2} \int d^4\theta X^\dagger X \chi^\dagger\chi
\label{nuevo}
\end{equation}
to masses 
\begin{equation}
m_{\Sigma^a}^2\simeq D^2/M^2\ .
\label{mSigma}
\end{equation}
If the $X$-superfields are located at an intersection of hidden stack
of branes separated from the observable sector three-brane along some
extra dimension, an operator similar to (\ref{nuevo}) does not appear
for fields in the observable sector and the corresponding transmission
of supersymmetry breaking is forbidden by locality or suppressed by
powers of the Planck scale $\sim F/M_{Pl}$. The actual value of the
$\Sigma$ mass is to a large extent model dependent but we will
implicitly assume that Eq.~(\ref{mSigma}) is at most suppressed by
one-loop factors just to cancel the negative contribution from
diagrams in Fig.~\ref{Sigmamass}.

The sector of quarks and leptons is made of $N=1$ chiral multiplets,
that we generically denote as $\mathcal Q=(Q,q_L)$. Its interactions
with the gauge sector are given by the Lagrangian
\begin{equation}
\int d^4\theta \mathcal Q^\dagger e^{V}\mathcal Q
\label{lag3}
\end{equation}
In principle, since the messenger sector has SM quantum numbers, there
are quartic interactions (from integration of the SM $D$-terms) as
\begin{equation}
\frac{1}{2}g_a^2\left(
\phi^\dagger_i\phi_j-\phi^c_i\phi^{c\dagger}_j\right)Q^\dagger_kQ_l
\sum_A(T^A_\phi)_{ij}(T^A_Q)_{kl}
\label{contacto}
\end{equation}
where $T^A_\phi$ are the generators of the gauge group in the
representation of $\phi$. From Eq.~(\ref{contacto}) and by performing
a one-loop contraction of the messenger fields $\phi$ and $\phi^c$,
one could in principle provide the SM sfermions $Q$ with a
phenomenologically unacceptable large squared mass $\sim D$. For
non-abelian group factors this contribution vanishes since it is
proportional to ${\rm Tr}(T^A_\phi)=0$. However, this cancellation
does not automatically take place for the case of $U(1)$ factors, as
the hypercharge. Moreover, the sign of the different sfermion squared
masses depends on their hypercharge and tachyonic masses can thus be
generated. A solution to this problem appears if $\phi$ is in a
complete representation of $SU(5)$ in which case it is guaranteed that
${\rm Tr}(Y_\phi)=0$~\footnote{Notice that this solution uses the fact
that within the states of the representation $\phi$ the mass matrix is
proportional to the unity and hence does not involve the vector-like
character of the hypermultiplet.}. This mechanism is similar to that
proposed in Ref.~\cite{Dimopoulos:1996ig} in conventional gauge
mediation. In fact, by making a $\pi/4$ rotation on the fields
$(\Phi,\Phi^{c\dagger})$, the mass matrix (\ref{mass0}) rotates to
\begin{equation}\left(
\begin{array}{cc}
M^2 & D\\
D& M^2
\end{array}
\right)
\label{mass1}
\end{equation}
which coincides with the supersymmetry breaking mass matrix in usual
gauge mediation via an $F$-term breaking when the messenger sector is
invariant under a ``messenger parity'' by just making the
identification $F=F^\dagger=D$.

In view of the previous identification and given that the adjoint
superfields $\Sigma$ do not have direct interactions with $N=1$
matter, the two-loop diagrams that contribute to the sfermions masses
are those in usual gauge mediation models (see the diagrams shown for
instance in Refs.~\cite{Alvarez-Gaume:1981wy,Poppitz:1996xw}) with the
result $m_Q^2\propto (\alpha/4\pi)^2 D^4/M^6$ where a strong
cancellation with respect to the usual gauge mediation of the
$F$-supersymmetry breaking due to the structure of the messenger mass
matrix is found~\footnote{We thank Yuri Shirman for a discussion on
this point.}. This contribution is however subleading since one-loop
diagrams involving Dirac gauginos in the effective theory where
messengers have been integrated out (three-loop diagrams in the
underlying theory) provide finite masses given
by~\cite{Antoniadis:1992eb,Fox:2002bu}
\begin{equation}
m_Q^2=\sum_{j=1}^3 k_jC_j(Q)\frac{\alpha_j}{\pi} (m_j^D)^2
\log\left[\frac{m_\Sigma^2}{(m_j^D)^2}\right]
\label{masaQ}
\end{equation}
where $C_j(N)=(N^2-1)/2N$ for the fundamental representation of
$SU(N)$.  

Finally associated with the superpotential term $\int d^2\theta
h_t\,\mathcal H^c \mathcal Q\,\mathcal U$ there exists the relevant
supersymmetry breaking parameter $A_t$ that appears in the
($R$-symmetry breaking) Lagrangian $-A_t h_t H^c Q\,U$. This term
generates a mixing between the left and right-handed stops and plays a
crucial role in the radiative corrections contributing to the lightest
Higgs mass. In our model we have $A_t=0$ at the scale $M$ and
furthermore, since the $D$ supersymmetry breaking does not break the
$R$-symmetry, this null value is not modified by the renormalization
group equations to any order in perturbation theory, unlike the case
of usual gauge mediation where the gaugino Majorana masses contribute
to the one-loop renormalization of $A_t$. However since, as we shall
see, $R$-symmetry is broken by tiny effects required for electroweak
symmetry breaking (and in particular also by gravitational
interactions) there will be a small trilinear parameter $A_t$ but
irrelevant for phenomenological purposes.

\section{\sc Electroweak breaking}
\label{electroweak}

Concerning the Higgs sector, it belongs to an $N=2$ hypermultiplet
$\mathbb H=(\mathcal H^c,\mathcal H)$ and its interactions with the
gauge sector are given by the Lagrangian
\begin{equation}
\int d^4\theta \left\{ \mathcal H^\dagger e^{V}\mathcal H+\mathcal H^c
e^{-V}\mathcal H^{c\dagger}\right\}-\left\{\sqrt{2}\,\int d^2\theta
\mathcal H^c \chi\mathcal H+h.c.\right\}
\label{lagH}
\end{equation}

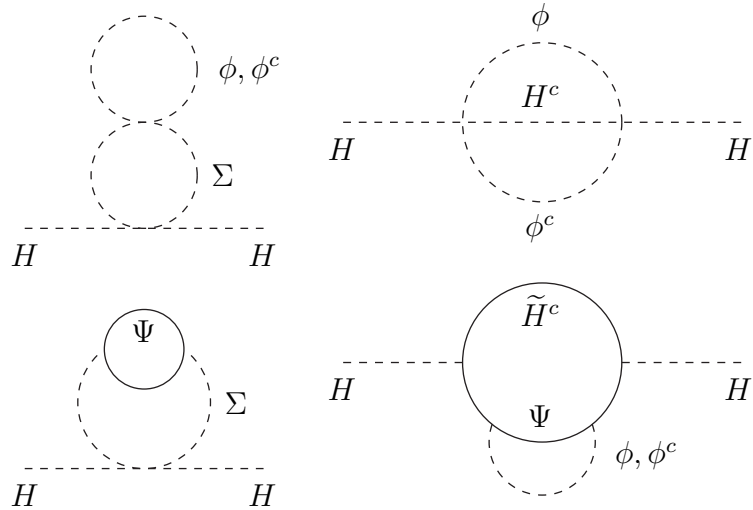
\begin{figure}[htb]
\begin{center}
\begin{picture}(300,90)(0,0)%\SetScale{2}
%\Vertex(180,10){1.5}\Vertex(120,10){1.5}
\DashLine(30,10)(120,10){3}
\DashCArc(75,30)(20,0,360){3}
\DashCArc(75,70)(20,0,360){3}
\Text(105,30)[]{$\Sigma$}
\Text(115,70)[]{$\phi,\phi^c$}
\Text(30,0)[]{$H$}\Text(120,0)[]{$H$}

%\Vertex(180,10){1.5}\Vertex(120,10){1.5}
\DashLine(150,50)(300,50){3}
\DashCArc(225,50)(30,0,360){3}
\Text(225,60)[]{$H^c$}
\Text(225,90)[]{$\phi$}
\Text(225,10)[]{$\phi^c$}
\Text(150,40)[]{$H$}\Text(300,40)[]{$H$}
\end{picture}
%\vspace{2cm}

\begin{picture}(300,90)(0,0)%\SetScale{2}
%\Vertex(180,10){1.5}\Vertex(120,10){1.5}
\DashLine(30,10)(120,10){3}
\DashCArc(75,35)(25,135,45){3}
\CArc(75,55)(15,0,360)

\Text(110,35)[]{$\Sigma$}
\Text(75,62)[]{$\Psi$}
\Text(30,0)[]{$H$}\Text(120,0)[]{$H$}

%\Vertex(180,10){1.5}\Vertex(120,10){1.5}
\DashLine(150,50)(195,50){3}\DashLine(255,50)(300,50){3}
\CArc(225,50)(30,0,360)
\DashCArc(225,20)(20,160,20){3}

\Text(225,70)[]{$\widetilde H^c$}
\Text(225,30)[]{$\Psi$}

\Text(265,15)[]{$\phi,\phi^c$}
%\Text(225,10)[]{$\phi^c$}
\Text(150,40)[]{$H$}\Text(300,40)[]{$H$}
\end{picture}

\caption{\it Feynman diagrams from $N=2$ superpotential contributing
to the $H$ squared mass. The same diagrams contribute also to the
$H^c$ squared mass (make the change $H\leftrightarrow H^c$ in all
graphs)}
\label{Higgsmass}
\end{center}
\end{figure}

The Higgs scalars acquire three-loop masses from the Feynman diagrams
leading to doublet slepton masses (\ref{masaQ}) and the two-loop
diagrams of Fig.~\ref{Higgsmass} that come from the $N=2$
superpotential gauge interactions in Eq.~(\ref{lagH}).

The contribution from the superpotential interactions is easily
written as
\begin{equation}
m_H^2=
2\sum_{j=1}^2 k_j C_j(H)\left(\frac{\alpha_j}{4\pi}\right)^2
\frac{D^2}{M^2}\left[1+\mathcal O\left(\frac{D^2}{M^4}\right)\right]
\label{masaH}
\end{equation}
that can be understood as coming from the effective operator
\begin{equation}
\sim\frac{1}{M^2}\int d^2\theta \mathcal W^2 \mathcal P_C
[\mathcal H^\dagger \mathcal H+\mathcal H^c \mathcal H^{c\,\dagger}]\, .
\end{equation}
where $\mathcal P_C$ is the non-local
operator~\cite{Gates:1983nr,Derendinger:1991hq}
\begin{equation}
\mathcal P_C=\Box^{-1}\bar{\mathcal D}\bar{\mathcal D}\mathcal D\mathcal D
\label{PC}
\end{equation}
that, acting on a real superfield, produces a chiral one such that its
lowest component contains the lowest component of the real superfield.
In our case the lowest component of ${\mathcal P_C} {\rm Tr}\mathcal
H^{\dagger}\mathcal H$ contains ${\rm Tr}H^{\dagger} H $.

Note that the Higgs potential is corrected with some additional
contributions coming from Eq.~(\ref{effec})~\cite{Fox:2002bu}.  The
resulting Lagrangian for the $\Sigma^A$ scalars, including the
radiative mass terms in Eq.~(\ref{mSigma}) can then be written as
\begin{equation}
-m_2^D D^a(\Sigma^a+\Sigma^{a\dagger})-m_1^D D_Y(\Sigma_Y+\Sigma_Y^{\dagger})
-m_{\Sigma^a}^2|\Sigma^a|^2-m_{\Sigma_Y}^2|\Sigma_Y|^2
\end{equation}
where $\Sigma^a$ and $D^a$ are the adjoint scalar and auxiliary field
of the $SU(2)$ vector superfield, $\Sigma_Y$ and $D_Y$ are the neutral
scalar and auxiliary field of the $U(1)$-hypercharge vector
superfield, and $m^D_{2}$ and $m^D_{1}$ represent the corresponding
gaugino Dirac masses.  We can now integrate out the adjoint fields
$\Sigma^a$ and $\Sigma_Y$. In the absence of the mass terms
(\ref{mSigma}) this integration would yield $D^a=D_Y=0$ which would be
a phenomenological drawback for this kind of models. However, in the
presence of the mass terms (\ref{mSigma}) the integration yields:
\begin{eqnarray}
\Sigma^a&=&-\frac{m_2^D \, D^a}{m^2_{\Sigma^a}} \nonumber\\
\Sigma_Y&=&-\frac{m_1^D\, D_Y}{m^2_{\Sigma_Y}}
\label{sigmavevs}
\end{eqnarray}
Replacing now (\ref{sigmavevs}) in the Lagrangian generates an extra
term (quartic in the Higgs fields)
\begin{equation}
\left(\frac{m_2^D}{m_{\vec\Sigma}}\right)^2\vec
D^2+\left(\frac{m_1^D}{m_{\Sigma_Y}}\right)^2D_Y^2
\end{equation}
which is a small correction to the tree-level Lagrangian
$-\frac{1}{2}(\vec D^2+D_Y^2)$ and can thus be neglected in the
calculation of the Higgs scalar potential. On the other hand, from
Eq.~(\ref{sigmavevs}), the VEV of the neutral component of the $SU(2)$
triplet $\Sigma^3$ gives rise to a small violation of the $SU(2)$
custodial symmetry when the neutral Higgses acquire a VEV. It would
contribute to the $\rho$-parameter as
\begin{equation}
\rho_0-1=4 \frac{\langle\Sigma^3\rangle^2}{v^2}=g^2\left(\frac{m_2^D
v}{m_{\Sigma}^2}\right)^2 \cos^22\beta
\label{rho}
\end{equation}
where $v=174$ GeV and as usual $\tan\beta$ is the ratio of the two
Higgs VEVs, $\tan\beta=v_2/v_1$.  Here we also used $\langle
D^3\rangle=1/2\ g v^2\cos 2\beta$, following from the usual
minimization of the Higgs potential.  Present experimental
bounds~\cite{Yao:2006px}, $\rho_0-1<6\times 10^{-4}$ are then easily
satisfied. Similarly, there is an induced VEV for the singlet
$\Sigma_Y$ that contributes to the $\mu$ parameter but without any
phenomenological impact.

Apart from the Higgs soft masses, there are two other key ingredients
of the Higgs sector in order to successfully break $SU(2)\times U(1)$:
namely the $\mu$ and $m_3^2$ terms. The former represents a
supersymmetric mass for Higgses and Higgsinos that can be written as
$\mu\int d^2\theta \mathcal H^c \mathcal H+h.c.$ Notice that this
superpotential mass term is perfectly consistent with the $N=2$
supersymmetry of the Higgs sector. However, its origin and correct
size is one of the main problems of supersymmetric model
building~\footnote{ Actually, as mentioned above, in our model such a
term is generated through the $N=2$ superpotential in
Eq.~(\ref{lagH}), upon the VEV of the scalar component $\Sigma_Y$ of
the superfield $\chi_Y$ in Eq.~(\ref{sigmavevs}). However its
magnitude is too low for phenomenological purposes and an additional
source of $\mu$ is needed.}. Moreover the $m_3^2$ term is a soft mass
for the scalar components of $\mathcal{H}$ and $\mathcal{H}^c$, as
$m_3^2 H^c H+h.c.$, which is required to explicitly break the
Peccei-Quinn symmetry.  The simplest possibility to generate a
$\mu$-term would be the existence in the $N=1$ observable sector of an
extra singlet chiral superfield $S$, coupled to the Higgs sector in
the superpotential $\int d^2\theta S\mathcal H^c\mathcal H$ and that
acquires a VEV $\langle S\rangle\sim $ TeV when supersymmetry is
broken, together with an $F$-component VEV generated (by some
O'Rafeartaigh mechanism or even by gravitational interactions as we
will see in the next section) at the scale $\sqrt{F_S}\sim $ TeV. This
breaking would be subleading with respect to the $D$-term breaking and
thus will not alter any of the general conclusions previously
obtained. Finally notice that, as we will comment in the next section,
$D$-breaking implies, in the presence of gravitational interactions
and through the cancellation of the cosmological constant, an
$F$-breaking that could be used to solve the $\mu$-problem by some of
the existing solutions in the literature either using
gravitational~\cite{Giudice:1988yz} or
gauge~\cite{Dvali:1996cu,Dine:1995ag} interactions.

The soft masses in the Higgs sector of Eq.~(\ref{masaH}) are `lowest
order' masses~\footnote{Both masses are equal because gauge
interactions respect the $N=2$ structure of the Higgs
hypermultiplet.}, that correspond to given boundary conditions at the
scale $M$, while the renormalization group equations (RGE) evolution
should be considered from $M$ to the weak scale $m_Z$. In particular,
the Higgs coupled to the top quark will get a large negative squared
mass proportional to the squared top-Yukawa coupling, as in the
MSSM~\footnote{Since the Yukawa couplings only respect $N=1$
supersymmetry we expect the renormalization of the two Higgs masses to
be different.}.  We can now write the Higgs potential in the usual
notation where $H_{1,2}$ are the lowest components of the chiral
superfields $\mathcal H$ and $\mathcal H^c$ respectively. For the
neutral components of the Higgs doublets the potential is~\footnote{We
are assuming here that the $\Sigma$ soft masses are generated (at
most) at one-loop order, as required by the cancellation of the
instability generated by the diagrams of Fig.~2, while $H_{1,2}$ soft
masses are generated at two-loop. In this way in the (effective)
potential (\ref{pot}) $\Sigma$-scalars are integrated out
non-supersymmetrically.}
\begin{eqnarray}
V&=&m_1^2 |H_1|^2+m_2^2 |H_2|^2-m_3^2(H_1 H_2+h.c.)\nonumber\\
&+&\frac{1}{8}(g^2+g'^2)(|H_1|^2-|H_2|^2)^2+
\frac{1}{2}(g^2+g'^2)|H_1H_2|^2
\label{pot}
\end{eqnarray}
where $m_{i}^2=m_{H_i}^2+\mu^2$ are the mass parameters at low energy
including the soft- and the $\mu$-terms. In the last line, the first
quartic term is the usual $D$-term of the MSSM, whereas the second is
a genuine $N=2$ effect similar to the one in Eq.~(\ref{lagH}).

Having an extra quartic term in the potential has interesting
consequences for its minimization.  Indeed, the origin in the Higgs
field space can either be a maximun or a saddle-point, whereas in the
MSSM it can only be a saddle-point due to the existence of a flat
direction of the $D$-term ($|H_1|=|H_2|$).  The conditions to have a
vacuum that breaks electroweak symmetry at the correct value are:
\begin{eqnarray}
&&\frac{m_Z^2}{2}=-\mu^2+\frac{1}{\tan^2\beta-1}\left(m_{H_1}^2-
m_{H_2}^2\tan^2\beta\right)\label{mZfix}\\ &&
m_A^2=m_{H_1}^2+m_{H_2}^2+2\mu^2+m_Z^2\label{mAfix}
\end{eqnarray}
where $m_A^2=2 m_3^2/\sin 2\beta$ is the squared mass of the
pseudo-scalar.  Notice the important fact that the MSSM stability
condition $2|m_3^2|<m_{H_1}^2+m_{H_2}^2+2\mu^2$ is not required in the
case of an $N=2$ Higgs sector. Actually the MSSM minimization
condition $m_A^2=m_{H_1}^2+m_{H_2}^2+2\mu^2$ is replaced by
Eq.~(\ref{mAfix}) so that for the same input mass parameters the value
of $m_A$ is larger than that of the MSSM. Another feature of the
potential is that the ``little'' electroweak fine-tuning of the MSSM
is reduced for values of $\tan\beta$ close to one, as we will see
below.

We can now calculate the spectrum of the Higgs sector. As in the MSSM,
it is controlled by $m_A$. However it has no dependence on
$\tan\beta$:
\begin{eqnarray}
m_h&=&m_Z\nonumber\\
m_H&=&m_A\nonumber\\
m^2_{H^\pm}&=&m^2_A+2\,m^2_W
\label{spec}
\end{eqnarray}
Moreover, the rotation matrix $R_\alpha$ from $H_1$, $H_2$ to $h$, $H$
is trivial leading to $\alpha=\beta-\pi/2$ which means that the
coupling $g_{Zhh}$ is the SM one, while $g_{ZHH}=0$; therefore $h$
behaves always like a SM Higgs and $H$ plays no role in electroweak
symmetry breaking. This leads to a different phenomenology from that
of the MSSM as we will describe in the next section.

The spectrum from Eq.~(\ref{spec}) is modified by radiative
corrections. At leading order, the mass matrix for the neutral states
can be written as:
\begin{equation}
\mathcal{M}=
\begin{pmatrix}
m_Z^2\cos^2\beta+m_A^2\sin^2\beta & (m^2_A-m^2_Z)\cos\beta\sin\beta \\
(m^2_A-m^2_Z)\cos\beta\sin\beta & m_Z^2\sin^2\beta+m_A^2\cos^2\beta+\epsilon
\end{pmatrix}
\end{equation}
where $\epsilon$ is the leading one loop correction which can be
written as:
\begin{equation}
  \epsilon=\frac{3 m_t^4}{4\pi^2v^2}\left(\log\frac{m^2_{\tilde{t}1}+
      m^2_{\tilde{t}2}}{m_t^2}+\frac{X_t^2}{2 M^2_S}
      \left(1-\frac{X_t^2}{6 M_S^2}\right)\right)
\end{equation}
where $m^2_{\tilde{t}1,2}$ are the two stop masses,
$X_t=A_t-\mu/\tan\beta\simeq -\mu/\tan\beta$ is the stop mixing mass
parameter, and $M_S\simeq\sqrt{(m_{\tilde t_1}^2+m_{\tilde t_2}^2)/2}$
represents the common soft supersymmetry breaking scale.  It should be
noted that for large $m_A$ and \emph{any value of $\tan\beta$} this
corresponds to the limit of the MSSM~\emph{at large $\tan\beta$} so
the Higgs mass in this model is always maximal in the decoupling limit
for the given value of $X_t$. In particular for values of
$\tan\beta\sim \mathcal O(1)$ we get a mixing $X_t^2\sim\mu^2$. On the
other hand, for values of $m_A$ close to $m_Z$ the value for $m_h$
should be below the present bound on the Higgs mass although such
small values should be normally excluded by the electroweak symmetry
breaking condition (\ref{mAfix}).

Finally it is easy to see that the ``little'' fine-tuning in this
model is greatly reduced with respect to that of the MSSM in the low
$\tan\beta$ region. The origin of the ``little'' fine-tuning in the
MSSM is that the Higgs mass only increases logarithmically with the
stop mass $m_Q$ while it appears quadratically in the determination of
$m_Z^2$. In fact since the tree-level mass of the Higgs in the MSSM
(in the large $m_A$ limit) goes to zero in the limit $\tan\beta\to 1$,
in order to cope with the LEP limit on the Higgs mass, one should go
to the region of very large values of $m_Q$ and thus to a very severe
``little'' fine-tuning. This fine-tuning is softened in the region of
large $\tan\beta$. Since in our model the tree-level Higgs mass does
not depend on $\tan\beta$ (it coincides with the MSSM one in the
$\tan\beta\to\infty$ limit) the ``little'' fine-tuning for any value
of $\tan\beta$ coincides with that of the MSSM in the large
$\tan\beta$ limit.

\section{\sc Some phenomenological aspects}
\label{phenomenology}

In this section we discuss different phenomenological aspects of the
model presented above.

\vspace{.5cm}\noindent{\large\sc The gravitino mass and gravitational
effects}

\noindent 
At the supergravity level, enforcing the cancellation of the
cosmological constant leads to the presence of an extra source of
supersymmetry breaking through an $F$-term for some chiral field. The
goldstino is then a combination of the fermionic partners of this
field and the fermionic partner of the $U(1)$ gauge boson with
non-vanishing $D$-term.  Through the super-Higgs mechanism it is
absorbed by the gravitino which gets a mass of
order~\cite{Cremmer:1984hj}
\begin{equation}
m_{3/2}\simeq  \frac{D}{M_{P\ell}}
\label{gravitino}
\end{equation}
where $M_{P\ell}=2.4\times 10^{18}$ GeV is the reduced Planck mass.
We will assume here that $F$ is of the same order as $D$:
\begin{equation}
F\simeq m_{3/2} M_{P\ell}\simeq D
\label{Fterm2}
\end{equation}
In fact, the relative sizes of $D$ and $F$ are very model dependent.
While normally $F\simgt D$, models with $F\ll D$ can also be
engineered~\cite{Gregoire:2005jr}.

The additional $F$-breaking source can be arranged to reside in the
hidden sector. It will be communicated only through gravitational
interactions to the observable sector. The associated effects
qualitatively differ from those from $D$-terms by the fact that they
could break $R$-symmetry, generate Majorana masses and are not
compelled to be flavor independent. This implies in particular that
they must be subleading with respect to the gauge mediated mechanism
presented in this paper in order to not re-introduce a flavor
problem. As the typical size of soft-mass terms induced by
gravitational interactions is $\sim F/M_{P\ell}$, if
Eq.~(\ref{Fterm2}) holds the condition for gravity mediated
contributions to be subleading is that $M\simlt
(\alpha/4\pi)\,M_{P\ell}\sim 10^{16}$ GeV which in turn implies the
bound $m_{3/2}\simlt~1$ TeV. However if $F\ll D$ then it may be
possible to suppress gravitational interactions even with
$m_{3/2}\simgt 1$ TeV.

Note also that anomaly-mediated supersymmetry breaking
(AMSB)~\cite{Randall:1998uk} is subleading with respect to the
soft-breaking induced by the gauge mediation mechanism, provided that
$M\ll M_{P\ell}$. For instance, the gaugino Majorana masses $M_a$
induced by anomaly mediation are
\begin{equation}
M_a\sim \frac{\alpha_a}{4\pi}\, m_{3/2}\sim
\left[\frac{M}{M_{P\ell}}\right] m_a^D
\end{equation}
and similarly for the squark and Higgs masses.

Fixing $m_a^D\sim 1$ TeV in Eq.~(\ref{mD}) one can write $D$ as a
function of $M$ as
\begin{equation}
D \sim M \times  10^5\, {\rm GeV}
\end{equation}
and plugging it into (\ref{gravitino}) one obtains
\begin{equation}
m_{3/2}\sim \left[\frac{M}{10^9 {\rm GeV}}\right]\ {\rm keV}
\end{equation}
Preferred ranges for values of the scales $M$ and $\sqrt{D}$ can then
be obtained if one imposes certain cosmological constraints on the
gravitino mass~\cite{Steffen:2006hw}. For instance as a warm dark
matter component a light gravitino mass should lie below $\sim 16$
eV~\cite{Viel:2005qj} in order not to have unwanted cosmological
consequences, which translates into $M\simlt 10^7$ GeV. Furthermore as
a cold dark matter component the mass of the gravitino should be above
few keV, which translates into $M\simgt 10^9$ GeV and the reheating
temperature has to be such that the density of gravitinos does not
overclose the universe~\cite{Viel:2005qj}.

\vspace{.5cm}\noindent{\large\sc Collider phenomenology}

\noindent One of the main differences between the usual gauge mediation and 
the models we have studied is the spectrum of supersymmetric particles
since in our case the scalars get masses at three-loops, in a very
similar way to theories with \emph{gaugino
mediation}~\cite{gaugino}. Although the precise spectrum depends on
the particular details on every model.

On the other hand there are three different experimental signatures of
these models which are quite general. The first one, and common to the
usual gauge mediation scenarios, arises when the gravitino is very
light (the LSP) and therefore the NLSP could have a decay suppressed
by the SUSY breaking scale. In this situation and if the NLSP is
charged (a very common case) it could yield a track in the detector
before decaying.

The second feature in our models is the $N=2$ structure of the Higgs
sector. As we noticed in section~\ref{electroweak}, the lightest Higgs
$h$ behaves as a SM Higgs with no $\tan\beta$ dependence on its
couplings to fermions; therefore its production and decay rates are
those typical of the SM rather than the MSSM. On the other hand, the
couplings to matter of the heaviest Higgs $H$, the pseudo-scalar $A$
and the charged Higgses $H^\pm$, as well as all self-couplings within
the Higgs sector do depend on $\tan\beta$: in principle one could then
distinguish these models from others (such as the MSSM or the
non-supersymmetric two-Higgs doublet models) by measuring these
couplings, although this may be the realm for ILC rather than LHC.

Finally the last signature comes from the $N=2$ structure of the gauge
sector and consequently from the fact that gauginos have their main
mass contribution forming a Dirac particle. This translates into
having different decay channels of gluinos, which are difficult to
measure but, more importantly, into the existence of \emph{three}
charginos and \emph{six} neutralinos, the latter being paired into
three pseudo-Dirac neutral fermions. Discovery of three charginos
should constitute the smoking gun of this scenario.

\vspace{.5cm}\noindent{\large\sc Unification}

\noindent Let us finish this phenomenological section with some
comments on gauge coupling unification. The model as it is, with $N=2$
gauge sector and a single Higgs hypermultiplet, together with the
usual $N=1$ chiral matter, has the following beta-function
coefficients:
$\beta_1=\frac{33}{5},\,\beta_2=3,\,\beta_3=0$. Evolution of the three
gauge couplings with the previous beta-functions does not lead to
unification. However, this can be achieved by adding appropriate extra
states~\cite{Kakushadze:1998vr,Fox:2002bu}. For instance, if one adds
to the above spectrum one hypermultiplet with the quantum numbers of
the Higgs plus two hypermultiplets with the quantum numbers of a
right-handed lepton (\emph{unifons}) the new beta-function
coefficients read:
\begin{equation}
\beta_1=\frac{48}{5},\quad\beta_2=4,\quad\beta_3=0
\end{equation}
\noindent
and one recovers the one-loop unification at the MSSM GUT scale
$M_{\rm GUT}\sim 2\times 10^{16}$ GeV. It should be noted that the
unification scale is not affected by the messengers since they form
complete $SU(5)$ representations.  Moreover, the \emph{unifons} come
in vector-like representations and can be given the desired
(supersymmetric) mass.

\section{\sc Conclusions}
\label{conclusion}
In this paper we have proposed a new model of gauge mediation where
supersymmetry is broken by a $D$-term expectation value in a {\it
secluded} local $U(1)$ sector, thus conserving $R$-symmetry in the
global limit. The NGMSB model, an alternative to the usual GMSB where
supersymmetry is broken by the $F$-term of a {\it secluded} chiral
sector, is easily realized in intersecting brane models of type I
string theory \cite{Antoniadis:2006eb}. Its main feature is the
presence of an extended $N=2$ supersymmetry in the gauge, as well as
in the Higgs and messengers sectors. As a result, the transmission of
supersymmetry breaking by the hypermultiplet messengers generates
Dirac masses for gauginos at the one-loop level. The observable matter
sector is contained in chiral $N=1$ multiplets (localized in brane
intersections) that get radiative masses at the one-loop order in the
effective theory where messengers have been integrated out. The NGMSB
model shares some features with the usual GMSB models:
\begin{itemize}
\item
It solves the supersymmetric flavor problem.
\item
It provides a common supersymmetry breaking scale in the observable
sector, the masses of all supersymmetric particles being proportional
to powers of gauge couplings.
\item
In sensible models, the gravitino is very light and thus a viable
candidate to describe the Dark Matter of the Universe.
\end{itemize}
However, NGMSB departures from usual GMSB in a number of
features:
\begin{itemize}
\item
Gauginos are the heaviest supersymmetric particles while squarks and
sleptons are one-loop suppressed with respect to them.
\item
The gaugino sector contains twice as many degrees of freedom as that
of the MSSM assembled into (quasi)-Dirac fermions with Dirac
masses. Finding three (instead of two) charginos, and six (instead of
four) neutralinos should then be the smoking gun of this class of
models at LHC.

\item
The two Higgs superfields of the MSSM are contained in one
hypermultiplet. Thus, the Higgs phenomenology for low $\tan\beta$ at
LHC is completely different from that of the MSSM. This alleviates the
fine-tuning problem of low $\tan\beta$ and opens up the corresponding
window in the supersymmetric parameter space.
\end{itemize}

There is a number of issues whose detailed analysis was outside the
scope of the present paper but that should be worth of further
study. On the phenomenological side, one should translate LEP data
into bounds on the Higgs masses~\footnote{Exclusion plots should look
very different from those in the MSSM.} and work out more in detail
the LHC phenomenology of both the scalar Higgs and the chargino and
neutralino sectors.  Finally at the string theory level, the
construction of realistic intersecting brane models realizing the
ideas contained in this paper.

\subsection*{\sc Acknowledgments}

\noindent 
Work supported in part by the European Commission under the European
Union through the Marie Curie Research and Training Networks ``Quest
for Unification" (MRTN-CT-2004-503369) and ``UniverseNet"
(MRTN-CT-2006-035863), in part by the INTAS contract 03-51-6346 and in
part by IN2P3-CICYT under contract Pth 03-1. The work of M.~Q. was
partly supported by CICYT, Spain, under contracts FPA 2004-02012 and
FPA 2005-02211. M.~Q. wishes to thank for hospitality the LPTHE of
``Universit\'e de Paris VI et Paris VII'', Paris, and the Galileo
Galilei Institute for Theoretical Physics, Arcetri, Florence, where
part of this work was done. The authors would like to thank Yuri
Shirman for pointing out several errors in the original version of
this paper.

\appendix
\section{\sc Tunneling probability density}
In this appendix we will first analyze the potential structure of the
$N=2$ supersymmetric $U(1)$ secluded sector $\mathbb A^0$ with gauge
coupling $g_0$ in the presence of the messenger hypermultiplet
$(\Phi,\Phi^c)$ with charges normalized to $\pm 1$. We will assume
that $N=2$ supersymmetry is broken by the FI parameter $\xi$. The
interaction of the messenger hypermultiplet $(\Phi,\Phi^c)$ with the
secluded $U(1)$ is described by the Lagrangian
\begin{equation}
\int d^4\theta \left\{ \Phi^\dagger e^{g_0V^0}\Phi+\Phi^c
e^{-g_0V^0}\Phi^{c\dagger}+\xi\,V^0\right\}+\left\{\int d^2\theta 
\Phi^c\left[M-\sqrt{2}\;g_0\chi^0 \right]\Phi+h.c.\right\}
\label{lag0}
\end{equation}
that gives rise to the scalar potential
\begin{equation}
V=\frac{D^2}{2g^2}+\left\{|M-\sqrt{2}\;\Sigma|^2+D\right\}|\phi^2|+
\left\{|M-\sqrt{2}\;\Sigma|^2-D\right\}|\phi^c|^2+
\frac{g^2}{2}\left(|\phi|^2+|\phi^c|^2\right)^2
\label{pot0}
\end{equation}
where we have used in (\ref{pot0}), to simplify the notation,
$D= g_0^2\xi$, $\Sigma= g_0\Sigma^0$ and $g= g_0$.

The potential (\ref{pot0}) has two minima:

$\bullet$ A local non-supersymmetric minimum at the origin,
$\phi=\phi^c=0$, which is a flat direction along the
$\Sigma$-field. This minimum has a vacuum energy $\langle
V\rangle=D^2/2g^2$. The flat direction is lifted by radiative
corrections and extra contributions that provide a mass to $\Sigma$ as
in Eq.~(\ref{mSigma})
\begin{equation}
\left.V_{\rm
rad}(\phi,\phi^c,\Sigma)\right|_{\phi=\phi^c=0}=m^2\,|\Sigma|^2,\quad
m^2\simeq \frac{1}{16\pi^2}\frac{D^2}{M^2}
\label{radmasa}
\end{equation}
Using Eq.~(\ref{radmasa}) the local supersymmetry breaking minimum is
then at the origin
\begin{equation}
\langle\phi\rangle=\langle\phi^c\rangle=\langle\Sigma\rangle=0
\label{localmin}
\end{equation}

$\bullet$ A global supersymmetric minimum at 
\begin{equation}
\langle\phi\rangle=0,\ |\langle\phi^c\rangle|^2=D/g^2,\
\langle {\rm Re} (\Sigma)\rangle=M/\sqrt{2},\langle{\rm Im}(\Sigma)\rangle=0
\label{globalmin}
\end{equation}
with zero vacuum energy~\footnote{The general features of our analysis
should remain true after introducing the interaction of the messengers
with the observable gauge sector. In particular, when the gauge sector
$\mathbb A^A$ is introduced, this minimum is uplifted and the
$\langle|\phi^c|^2\rangle$ VEV is shifted by the observable $D$-term
$g_A^2/2\left(\phi^c T^A\phi^{c\,\dagger}\right)^2$, while the
$\Sigma$ direction should have components along the $\Sigma^A$-fields
of the observable sector. However radiative corrections (\ref{mSigma})
provide masses proportional to ${\rm tr} T_{\phi}^A T_{\phi}^B+{\rm
tr} T_{\phi^c}^A T_{\phi^c}^B$. Given that
$T^{U(1)}_{\phi,\phi^c}\propto 1$ and ${\rm tr} T^{Y}_{\phi,\phi^c}=0$
there is no mixing between the {\it secluded} $U(1)$ and $U(1)_Y$ and
all flat directions are lifted.  Since we intend to do only a
semi-quantitative analysis of the tunneling probability from the local
(\ref{localmin}) to the global (\ref{globalmin}) minimum we do not
include in this appendix gauge interactions from the observable
sector.}.

For any value of $\Sigma$ at zero mass, the potential has a minimum at
the origin along the $\phi$-direction and for values of $\Sigma$ such
that $|M-\sqrt{2}\,\Sigma|^2>D$ the potential also has the minimum
(\ref{globalmin}) along the $\phi^c$ direction. However when
$|M-\sqrt{2}\,\Sigma|^2=D$ there is an almost flat direction along
$\phi^c$ at the minimum that becomes a maximum for
$|M-\sqrt{2}\,\Sigma|^2<D$. Since the flat direction $\Sigma$ is
lifted only by radiative corrections [see Eq.(\ref{radmasa})] the path
followed by the instanton which controls the tunneling from the local
(\ref{localmin}) to the global (\ref{globalmin}) minimum goes along
the ${\rm Re}(\Sigma)$-direction, from the origin up to values of
${\rm Re}(\Sigma)$, ${\rm Re}(\Sigma)=M/\sqrt{2}+\mathcal O(D^{1/2})$
where the potential becomes unstable along the $\phi^c$
direction. Since the instanton has to jump over the path of least
slope along ${\rm Re}(\Sigma)$, up to values of ${\rm Re}(\Sigma)$
where the potential becomes unstable, the tunneling problem can be
very well approximated by a one-dimensional problem where the
instanton field that satisfies the euclidean equation of
motion~\cite{Coleman:1977py}
\begin{equation}
\frac{d^2\varphi}{dr^2}+\frac{3}{r}\frac{d\varphi}{dr}=V^\prime(\varphi)
\label{insteq}
\end{equation}
is $\varphi\equiv {\rm Re}(\Sigma)/\sqrt{2}$.  The variable
$r=\sqrt{t^2+\vec x^2}$ in (\ref{insteq}) makes the $O(4)$ symmetry of
the solution manifest with the boundary conditions $\varphi\to 0$ at
$r\to\infty$ and $d\varphi/dr=0$ at $r=0$. 

In order to make an estimate of the probability of bubble formation by
tunneling from the local (\ref{localmin}) to the global
(\ref{globalmin}) we will follow Ref.~\cite{Linde:1981zj}. First of
all notice that the depth of the global minimum $\sim D^2/2g^2$ is
much larger than the heigh of the barrier $\sim D^2/32\pi^2$ so that
the bubble solution is outside the domain of validity of the thin wall
approximation. Second, to compute the tunneling probability one can
disregard the details of the behaviour of the potential $V(\varphi)$
at $\varphi\gg\varphi_1$ where $V(\varphi_1)\simeq V(0)$. In
particular considering a potential $V_{\rm app}(\varphi) $ that
approximates $V(\varphi)$ for $\varphi\simlt \varphi_1$ is usually a
sufficiently good approximation~\cite{Linde:1981zj}. In our case
$\varphi_1\simeq M/2$ and the simplest such potential is
\begin{equation}
V_{\rm
app}(\varphi)=\frac{1}{2}m^2\,\varphi^2-\frac{\delta}{3}\,\varphi^3
\label{apppot}
\end{equation}
where the mass term is given by Eq.~(\ref{radmasa}) and $\delta$ is
chosen such that $V(\varphi_1)\simeq 0$, i.~e.
\begin{equation}
\delta\simeq \frac{3m^2}{2\varphi_1}\sim
\frac{3}{16\pi^2}\frac{D^2}{M^3}
\label{quartic}
\end{equation}
The euclidean action $S_4$ corresponding to the solution $\varphi$ of
Eq.~(\ref{insteq}) is given by
\begin{equation}
S_4%\sim2\left(\frac{10\, m}{\delta} \right)^2 
\sim2\left(\frac{10 M}{3\,
m} \right)^2 \sim \left(\kappa_4\, \frac{M^2}{D} \right)^2
\label{s4}
\end{equation}
where $\kappa_4\simeq 59$.

At finite temperature one should replace $V(\varphi)$ by
$V(\varphi,T)$ where finite temperature effects have been
considered. In our case the dominant thermal effects in
Eq.~(\ref{apppot}) can be encoded in the thermal mass
\begin{equation}
m^2(T)\simeq \frac{1}{16\pi^2}\left(\frac{D^2}{M^2}+a T^2\right)
\end{equation}
where $a$ is some order one coefficient. At finite temperature the
$O(4)$ symmetric solution of (\ref{insteq}) should be replaced by the
$O(3)$ symmetric one that satisfies the equation~\cite{Linde:1981zj}
\begin{equation}
\frac{d^2\varphi}{dr^2}+\frac{2}{r}\frac{d\varphi}{dr}=V^\prime(\varphi,T)
\label{insteqT}
\end{equation}
where now $r^2=\vec x^2$. For the potential (\ref{apppot}) it is
found~\cite{Linde:1981zj}
\begin{equation}
\frac{S_3}{T}%\sim \frac{44\, m^3(T)}{\delta^2 T}
\sim 5\,\frac{M^3
m^3(T)}{m^4\,T} \simgt\left(\kappa_3\,\frac{M^2}{D}\right)^2
\label{s3}
\end{equation}
where $\kappa_3\simeq 45$ and the last inequality holds for any
temperature.
The tunneling probability per unit time per unit volume is then
\begin{equation}
\mathcal P\sim e^{-\mathcal B}
\label{B}
\end{equation}
where $\mathcal B=\min\left[S_4,S_3(T)/T \right]$ at any
temperature. This justifies our Eq.~(\ref{tunnel}).

Of course we do expect expressions (\ref{s4}) and (\ref{s3}), based on
the approximated potential (\ref{apppot}) to be accurate only within
factors of order a few. In fact if we approximate the potential by a
different one with a negative quartic term, $V_{\rm app}=1/2\,
m^2\varphi^2-1/4\,\lambda\varphi^4$ as in Ref.~\cite{Linde:1981zj},
and fix $\lambda$ such that $V_{\rm app}(\varphi_1)\simeq 0$ the
corresponding expressions (\ref{s4}) and (\ref{s3}) have coefficients
$\kappa_4\simeq 23$ and $\kappa_3\simeq 19$ respectively, thus related
to those in (\ref{s4}) and (\ref{s3}) by order two factors. Given that
in all cases the euclidean actions are dominated by the large
$M^4/D^2$-factors, these results are very consistent to each other.

\end{document}